\documentstyle[amsfonts,amscd,esfconf]{amsart}

\begin{document}
\title{Wick Type Symbol and Deformed Algebra of Exterior Forms}
\authors{V. A. Dolgushev\adref{1},
S.L. Lyakhovich\adref{1} and A.A. Sharapov\adref{1}}
\addresses{\1ad Department
of Physics, Tomsk State University}
\maketitle

\begin{abstract}
The covariant descripion is constructed for the Wick-type symbols
on symplectic manifolds by means of the Fedosov procedure. The geometry of
the manifolds admitting this symbol is explored.
The superextended version of the Wick-type
star-product is introduced and a possible application of the
construction to the noncommutative field theory is discussed.
\end{abstract}






\section{Introduction}

The aim of the paper is twofold. First we  construct
covariant  Wick-type star-product within the
framework of the Fedosov deformation quantization \cite{Fedosov}. Second we
propose a canonical superextension of the construction along the lines of
the Bordemann approach \cite{SBor} and present a Wick-type deformation for
the exterior algebra of the base even manifold. The first step is
motivated by a wide physical application of the Wick-type star-product
especially for the models with the infinite degrees of freedom.
The second one is motivated by noncommutative field theory
\cite{CR,CDS,SW} on a curved manifold, where the covariant definition
 is required for the
deformed product between the {\it tensor observables}.
We also hope that recently found relationship between the Fedosov deformation
and BRST theory \cite{GL}, being combined to the Wick type symbol technique,
should pave the way to studying anomalies in a wide class of the filed theory
problems making use of the deformation quantisation tools.

\section{The Wick-type symbol in the deformation quantization}

Under the term ``Wick-type'' we understand a broad
class of symbols incorporating, along with the ordinary (genuine) Wick
symbols, the so-called $qp$-symbols as well as various mixed possibilities
commonly regarded as the pseudo-Wick symbols. To give a more precise
definition of what is meant here consider first the linear symplectic
manifold ${\Bbb R}^{2n}$ equipped with the canonical Poisson brackets $%
\{y^i,y^j\}=\omega ^{ij}$. Then the usual Weyl-Moyal product of two
observables, defined as
$$
a*b(y)=\exp \left( \frac{i\hbar }2\omega ^{ij}\frac \partial {\partial
y^i}\frac \partial {\partial z^j}\right) a(y)b(z)|_{z=y},
$$
turns the space of smooth functions in $y$ to the non-commutative
associative algebra with unit called the algebra of Weyl symbols. The
transition from the Weyl to Wick-type symbols is achieved by adding
a certain complex-valued symmetric tensor $g$ to the Poisson one $\omega $
in the formula for the Weyl-Moyal $*-$product,
\begin{equation}
a*_gb(y) =\exp \left(
\frac{i\hbar }2\Lambda ^{ij}\frac \partial {\partial y^i}\frac \partial
{\partial z^j}\right) a(y)b(z)|_{z=y}, \label{wlp}
\end{equation}
\begin{equation*}
\Lambda ^{ij}
=\omega ^{ij}+g^{ij},\qquad
\Lambda ^{\dagger }=-\Lambda .
\end{equation*}
Although the associativity of the modified product holds for any constant $g$
the Wick type symbols are extracted by the additional half-rank condition $%
rank\Lambda =n$. In particular, the genuine Wick symbol corresponds to a
pure imaginary $g$ while the real $\Lambda $ is associated with $qp$-symbol.
In the general case one can easily show that the dual complexified phase
space ${\Bbb C}^{2n}$ is splitted into a direct sum of two transverse
Lagrangean subspaces, which are left and right kernel subspaces for the
matrix $\Lambda ^{ij}$\thinspace .

Turning to the curved manifold $M$ we just replace the constant matrix $%
\Lambda ^{ij}$ by a general complex-valued bilinear form $\Lambda
^{ij}(x)=\omega ^{ij}(x)+g^{ij}(x)$ with the antisymmetrical part being the
non-degenerate Poisson tensor. Then the Wick-type star-product is defined to
satisfy the following ``boundary condition''
\begin{equation}
\label{bound}a*b=ab-\frac{i\hbar }2\Lambda ^{ij}\partial _ia\partial
_jb+\ldots
\end{equation}
$\hbar $ being the formal deformation parameter (``Plank constant''), and
dots mean the terms of higher orders in $\hbar .$ As in a general case of
the symplectic manifold the condition (\ref{bound}) is compatible with
so-called the {\it correspondence principle }of quantum mechanics:
\begin{equation}
\label{bb2}\lim _{\hbar \rightarrow 0}\frac i\hbar (a*b-b*a)=\{a,b\},
\end{equation}
where $\{\cdot ,\cdot \}$ means the Poisson bracket associated to $\omega
^{ij}$. It turns out that if there exists a torsion-free linear connection $%
\nabla $ preserving $\Lambda ^{ij}$ the star-product satisfying the
condition (\ref{bound}) can be achieved by a minimal modification of the
Fedosov method. Namely, if we replace the $\circ $-multiplication in the
Weyl algebra bundle \cite{Fedosov} by the that of the Wick type (\ref{wlp})
all the steps and the theorems of the Fedosov method is generalized in a
straightforward manner and thus we get the star-product of the Wick type.

The only crucial point of this program is the existence of a torsion-free
linear connection $\nabla $ preserving $\Lambda $. One may readily find that
the necessary and sufficient condition for such a connection to exist is the
integrability of the right and left kernel distribution of $\Lambda (x).$
When the latter condition is fulfilled $\nabla $ is just the Levi-Civita
connection associated to the symmetric and non-degenerate form $g(x)$ and
the right and left kernel distributions define the transverse polarizations
of the symplectic manifold $(M,\omega )$. Thus we see that the symplectic
manifold admitting the Wick-type star-product are necessarily equipped with
the pair of the transverse polarizations.

After the paper \cite{GRSh}, the symplectic manifold equipped by a
torsion-free symplectic connection is usually called as Fedosov manifold
because precisely these data -- symplectic structure and connection -- enter
to Fedosov's star-product. The Wick deformation quantization (as we have
defined) involves one more geometric structure - a pair of transverse
polarizations, and, by analogy to the previous case, the underlying manifold
may be called as the \textit{Fedosov
manifolds of Wick type} or \textit{FW-manifold for short}. The extended list
of examples of the FW-manifolds is provided by the K\"ahler manifolds. It is
the case when the matrix of the tensor $\Lambda ^{ij}$ is the anti-Hermitian
in local real coordinates. There are also examples of the FW-manifolds
having no K\"ahler structure but assigned instead by two real
transverse polarizations.
The last situation is strikingly illustrated
by the one-sheet hyperboloid embedded into three dimensional Minkowski space
as the surface
\begin{equation*}
x^2+y^2-z^2=1
\end{equation*}
$x,y,z$ being linear coordinates in $R^{2,1}$.
one ($\det g<0$). The integral leaves of the corresponding real polarizations
coincide here with two transverse sets of linear generatings
of hyperboloid's surface being, in turn, the isotropic geodesics of the
respective metric structure.

At present there is a large amount of literature concerning the
deformation quantization on polarized symplectic manifolds
(see i.g. \cite{Mol},\cite
{Bor1},\cite{Kar} and references therein) beginning with the pioneering paper
by Berezin \cite{Ber} on the quantization in complex symmetric spaces.
However, in all the papers cited above the very definition and the
construction of the Wick-type star-product are based on the explicit use of
local coordinates adapted to the polarization (or \textit{separation of
variables} in terminology of work \cite{Kar}).

\section{Superextension}

Now we present construction of the double dimensional superextension of the
FW-manifold $M$ and perform its deformation quantization. We construct the
supermanifold supplying the initial FW-manifold by additional odd variables $%
\theta ^i$, where $i$ is the index of the tangent space. Then the formal
series

\begin{equation}
{\cal C}\ni a=a(x,\theta )=\sum\limits_{k=1}^{2^{2n}}a(x)_{i_1\ldots
i_k}\theta ^{i_1}\ldots \theta ^{i_k},
\label{sfunc}
\end{equation}
where the coefficients $a(x)_{i_1\ldots i_k}$ are antisymmetric with respect
to the index permutations are thought to be superfunctions constituting a
supercommutative superalgebra, which we will denote by ${\cal C}$ . As it
follows from the results by Bordemann \cite{SBor} the Grassmann
multiplication in ${\cal C}$ can be formally deformed into associative $%
{\Bbb {Z}}_2$-graded multiplication $*$ on ${\cal C}[[\hbar ]]$ whenever the
symplectic and the metric structures are given on the base even manifold.
The remarkable feature of this construction is that the deformation
quantization for the algebra of super functions is performed first and the
super Poisson bracket arises here {\it a posteriori }as $\hbar $-linear term
of the supercommutator.

Here we modify Bordemann's approach to the quantization of the supermanifold
\cite{SBor} and present the deformation quantization of the Wick type for
the superextension of the FW-manifold. Since the basic steps of our
construction are essentially similar to those in the original
papers by Fedosov
\cite{Fedosov} and
Bordemann \cite{SBor} we
consider them very briefly dwelling only upon peculiarities.

First we define a superextension of the Weyl algebra bundle ${\cal W}$,
\thinspace introduced in \cite{Fedosov}\-.

\vspace{0.5cm} \noindent
{\bf Definition 1. }{\it The bi-graded superalgebra }${\cal SA}=\oplus
_{m,n=0}^\infty ({\cal SA})_{m,n}${\it \ with unit over }${\Bbb {C}}$ is{\it %
\ a space of formal series, }
\begin{equation}
\label{sect}a(x,\theta ,y,dx,\hbar )=\sum_{2k+p\ge 0}\hbar ^ka_{k\,i_1\ldots
i_pj_1\ldots j_{q^{\prime }}l_1\ldots l_{q^{\prime \prime
}}}(x)y^{i_1}\ldots y^{i_p}\theta ^{j_1}\ldots \theta ^{j_{q^{\prime
}}}dx^{l_1}\ldots dx^{l_{q^{\prime \prime }}},
\end{equation}
{\it multiplied with the help of associative }$\circ ${\it -product of the
form}
\begin{equation}
a\circ b=
\end{equation}
\begin{equation*}
=exp\frac{i\hbar }2\Lambda ^{ij}(x)\left( \frac \partial {\partial y^i}\frac
\partial {\partial z^j}+\frac{\overset{\leftarrow }{\partial }}{\partial
\theta ^i}\frac{\overset{\rightarrow }{\partial }}{\partial \chi ^j}\right)
a(x,y,\theta ,dx,\hbar )b(x,z,\chi ,dx,\hbar )|_{y=z,\theta =\chi }
\end{equation*}
{\it A general term of the series (\ref{sect}) is assigned by the bi-degree }%
$(2k+p+q^{\prime },q^{\prime \prime })${\it .}

The expansion coefficients $a_{k\,i_1\ldots i_pj_1\ldots j_{q^{\prime
}}l_1\ldots l_{q^{\prime \prime }}}(x)$ are considered to be components of
covariant tensors on $M$ symmetric in $i_1,\ldots ,i_p$ and antisymmetric
with respect to $j_1\,,\ldots \,,j_{q^{\prime }}$ and $l_1\,,\ldots
\,,l_{q^{\prime \prime }}$\thinspace . Note that we omit the sign $\wedge $
between the differentials regarding $dx^i$ and $\theta ^j$ as the set of $2n$
anti-commuting variables:
\begin{equation}
dx^idx^j=-dx^jdx^i,\quad \theta ^i\theta ^j=-\theta ^j\theta ^i,\quad
dx^i\theta ^j=-\theta ^jdx^i
\end{equation}
Hence, ${\cal SA}$ is also ${\Bbb {Z}}_2$-graded algebra with respect to the
Grassmann parity $q=q^{\prime }+q^{\prime \prime }$ (\ref{sect}) and the
supercommutator of two homogeneous elements $a,b\in {\cal SA}$ with the
parities $q_1$and $q_2$ is defined as
\begin{equation}
[a,b]=a\circ b-(-1)^{q_1q_2}b\circ a
\end{equation}
Let us introduce the nilpotent operator
$\delta =dx^i\frac \partial {\partial y^i}$ which will be rather
important in formulating the respective analogues of the Fedosov
equation. It is easily seen to be the
inner superderivation of the superalgebra ${\cal SA}$\,.

The nontrivial cohomology of $\delta$
coincides with the space of superobservables ${\cal C}[[\hbar ]]$
and the operator $\delta ^{-1}$
\begin{equation}
\label{b9}\delta ^{-1}a=y^ki\left( \frac \partial {\partial x^k}\right)
\int\limits_0^1a(x,ty,tdx,\hbar )\frac{dt}t,
\end{equation}
is the partial homotopy operator for $\delta $ in the sense of ``Hodge-De
Rham'' decomposition

\begin{equation}
\label{b10}a=\sigma (a)+\delta \delta ^{-1}a+\delta ^{-1}\delta a\,,
\end{equation}
where $\sigma (a)=a(x,\theta ,0,\hbar )$ and the interior derivation
$i\left( \frac \partial {\partial x^k}\right)$ acts on the forms by the
rule
$$
i\left( \frac \partial {\partial x^k}\right) a_{i_1\ldots i_m}
dx^{i_1}\ldots dx^{i_m}
= m a_{ki_2 \ldots i_m} dx^{i_2}\ldots dx^{i_m}
$$

The covariant derivative $\nabla $ given on the base FW-manifold induces the
superderivation in ${\cal SA}$
\begin{equation}
\label{na}\nabla :({\cal SA})_{n,m}\rightarrow ({\cal SA})_{n,m+1},
\end{equation}
\begin{equation*}
\nabla =dx^i\left( \frac \partial {\partial x^i}-y^j\Gamma _{ij}^k(x)\frac
\partial {\partial y^k}-\theta ^j\Gamma _{ij}^k(x)\frac{\overset{\rightarrow
}{\partial }}{\partial \theta ^k}\right) ,
\end{equation*}
$\Gamma _{ij}^k$ are Christoffel symbols of the connection associated to $%
\Lambda $. It is easy to check that $\nabla $ anti-commutes with $\delta $
and its curvature is given by

\begin{equation}
\label{n^2}\nabla ^2a=\nabla (\nabla a)=\frac 1{i\hbar }[R\,,\,a],
\end{equation}
\begin{equation}
\label{sr}R=\frac 14R_{ijkl}y^iy^jdx^kdx^l+\frac 14{\cal R}_{ijkl}\theta
^i\theta ^jdx^kdx^l
\end{equation}
where we have used the notations $R_{ijkl}=\omega _{im}R_{~jkl}^m$ and $%
{\cal R}_{ijkl}=g_{im}R_{~jkl}^m$. Following Fedosov method, one can
combine $\nabla $ with an inner derivative to get the Abelian connection of
the form

\begin{equation}
\label{asd}D=\nabla -\delta +\frac 1{i\hbar }[r,\cdot ]=\nabla +\frac
1{i\hbar }[\omega _{ij}y^idx^j+r,\cdot ],\quad r=r_i(x,\theta \,,y,\hbar
)dx^i,
\end{equation}
Denote ${\cal SA}_D=\ker D$. The next assertions is the super counterpart of
that stated in Fedosov's original theorems \cite[Theorems 3.2, 3.3]{Fedosov}.

\noindent
{\bf Theorem 1. }{\it With the above definitions and notations we have:}

{\it i) there is a unique Abelian connection }$D${\it \ (\ref{asd}) for
which }
\begin{equation*}
\delta ^{-1}r=0,\quad r_i(x,\theta \,,0,\hbar )=0
\end{equation*}

{\it and }$r${\it \ consists of monomials whose first degree is no less then
3;}

{\it ii) }${\cal SA}_D${\it \ is a subalgebra of }${\cal SA}${\it \ and the
map }$\sigma ${\it \ being restricted to }${\cal SA}_D$

{\it defines a linear bijection onto }${\cal C}[[\hbar ]]${\it ;}

\vspace{0.5cm}\noindent
{\bf Corollary 1}. {\it The pull-back of }$\circ ${\it -product via }$\sigma
^{-1}${\it \ induce an associative }$*${\it -product on the space of
physical observables }${\cal C}[[\hbar ]],${\it \ namely}
\begin{equation}
\label{b16}a*b=\sigma (\sigma ^{-1}(a)\circ \sigma ^{-1}(b))
\end{equation}

\vspace{0.5cm} \noindent
{\bf Proof }may be directly read off from \cite[Theorems 2.1,2.2]{SBor}.

The peculiar property of the presented Wick-type deformation is that the
superalgebras $({\cal C}[[\hbar ]], *)$ and $(C^{\infty}(M)[[\hbar]],*')$\,,
where $*'$ is the respective Wick-type star-product on the base (even)
manifold are related to each other in the sense of the following proposition

\vspace{0.1cm} \noindent
{\bf Proposition 1.} {\it Let} $\pi :{\cal C}[[\hbar ]]\rightarrow C^\infty
(M)[[\hbar ]]${\it \ be the canonical projection defined as} $\pi
a(x,y,\hbar ,\theta )=a(x\,,y\,,\hbar ,0)$. {\it Consider the algebras} $%
{\cal C}[[\hbar ]]${\it \ and }$C^\infty (M)[[\hbar ]]${\it \ as left(right)
moduli over }$C^\infty (M)[[\hbar ]]${\it . Then }$\pi ${\it \ defines a
homomorphism of the introduced moduli, that is for }$\forall a\in ${\it \ $%
C^\infty (M)[[\hbar ]]$\ and }$\forall b\in {\cal C}[[\hbar ]]${\it \ \ we
have:}

\begin{equation}
\label{induce}a*^{\prime }(\pi b)=\pi (a*b),\quad (\pi b)*^{\prime }a=\pi
(b*a)
\end{equation}
It is easy to check that the equations (\ref{induce}) do not hold for the
case of the Weyl star-product obtained for example by Bordemann's approach
\cite{SBor}.

\section{Conclusion}

In this paper we have realized the Wick-type star-product following to the
Fedosov approach to the deformation quantization. We have observed that the
most natural geometrical framework for this star-product is the so-called
Fedosov manifold of Wick type. This is defined as the real symplectic
manifold equipped with a complex-valued half-rank bilinear form $\Lambda $
preserved by some torsion-free connection and it is shown to be necessarily
the manifold admitting two transverse polarizations.

We have shown that any FW-manifold can be canonically extended to the
doubled dimensional supersymplectic manifold associated to the tangent
bundle of the initial manifold. Moreover, the latter can be easily then
quantized along the line of the Fedosov approach and this quantization
procedure, yields the natural definition of the deformed product between
antisymmetrical covariant tensors on the base even manifold. The latter is
due to the natural one-to-one correspondence between the elements of
the exterior algebra and the superfunctions (\ref{sfunc}) of the introduced
supermanifold. This correspondence can be easily restored by its
definition on the
monomial elements of the exterior algebra
$$
\omega_{i_1 \ldots i_p}dx^{i_1}\ldots dx^{i_p} \mapsto
\omega=\omega_{i_1 \ldots i_p}\theta^{i_1}\ldots \theta^{i_p}
$$

We hope that this deformed associative product may
find its application in the
noncommutative field theory in particular for constructing the
noncommutative Yang-Mills theory \cite{CR,CDS,SW} on the curved symplectic
manifolds. The problem of realizing the
noncommutative Yang-Mills theory on a curved manifold
is firstly posed in the paper by Connes, Douglas and Schwarz \cite{CDS}
and at first sight it seems to
have a natural
solution in the context of the deformation quantization. However the
deformation quantization poses the problem of constructing the deformed
algebra of scalar functions on the symplectic manifold. Hence it is
insufficient for constructing the noncommutative Yang-Mills theory, which
should describe the dynamics of tensor fields as well. Thus one firstly has
to realize the star-product between tensor observables on the symplectic
manifold. The star-product in addition should be defined in a generally
covariant way since otherwise it would be impossible to present the
invariant action for the theory in question. The quantization of the
supermanifold presented above suggests a possible solution for the problem
as it yields the definition for the covariant deformation of the exterior
algebra on the initial even manifold. However the deformed algebra of the
forms cannot be restricted for example to the set of the 1-forms only thus a
hypothetical noncommutative Yang-Mills theory would describe the dynamics of
the forms of all the possible ranks.

Another ingredient which is important for constructing the action for
the noncommutative Yang-Mills
theory is the metric structure. This structure, in turn, being
naively inserted into the action functional may violate the gauge invariance
of the theory even in the case of the ordinary Weyl-Moyal star-product.
Thus the metric and symplectic structures should be considered
on equal foot just
like in the construction of the Wick-type star-product. Finally the
construction of the gauge invariant action requires the
definition of the trace measure for the aforementioned superalgebra.
We will consider this problem elsewhere.

\noindent{\bf Acknowledgments.} It is our pleasure to express our sincere
gratitude to
I. Batalin, I. Gorbunov, M. Grigoriev,  M. Henneaux
A. Karabegov, R.Marnelius, A. Nersessian and M.Vasiliev
for useful discussions related to various aspects of this paper.
This
work has been partially supported by RFBR grant \symbol{242} 00-02-17-956.


\end{document}